\documentclass[preprint,showpacs,preprintnumbers,amsmath,amssymb,superscriptaddress]{revtex4}

\usepackage[shortlabels]{enumitem}
\usepackage{graphicx,amsfonts}
\usepackage{physics,slashed}

\begin{document}

\title{Electric charge assignment in quantum field theories}

\author{E. Castillo-Ruiz}
\email{ocastillo@uni.edu.pe}
\affiliation{Facultad de Ciencias, Universidad Nacional de Ingenier\'\i a, Lima Per\'u}

\author{V. Pleitez}
\email{v.pleitez@unesp.br}
\affiliation{Instituto de F\'\i sica Teórica - Universidade Estadual Paulista, São Paulo, Brazil.}

\author{O. P. Ravinez }
\email{opereyra@uni.edu.pe}
\affiliation{Facultad de Ciencias, Universidad Nacional de Ingenier\'\i a, Lima Per\'u}
\affiliation{
ICTP South American Institute for Fundamental Research\\ Instituto de F\'\i sica Te\'orica Universidade--Estadual Paulista\\ Rua Dr. Bento T. Ferraz 271, 01140-70, São Paulo, Brazil.}

\date{09/01/2020}

\begin{abstract}

In Field theories with simple or semi-simple unitary, local or global symmetries, the electric charge is related to a global one. This is the case also in electroweak gauge theories even before the spontaneous symmetry breaking (SSB), where these quantities are defined in order to expand the Lagrangian respecting the conservation laws after the breaking. So, the electric charge assignment within a given multiplet is done (in units of $\abs{e}$) considering only global symmetries, even though in general, the electric charge operator is a linear combination of the diagonal generators of the non-Abelian symmetry plus some Abelian factor. In this work we show how this operator can be systematically constructed from any representation in the Standard Model and some of its extensions as the 3-3-1, left-right symmetric and grand unified models.
\end{abstract}


\maketitle

\section{Introduction}

The assignment of electric charges to the particle fields is a task already verified for the case of global unitary symmetries like the hadron $SU(2)_V$ or $SU(3)$, just to cite the two main cases. Additionally in models with spontaneous symmetry breaking (SSB), the electric charge assignment does not distinguish if the main symmetry is global or local because the electric charge is a quantum number that is conserved globally although the Lagrangian belongs to a local $U(1)$ symmetry after SSB.

For instance, in Standard Model (SM) symmetry, all particles have no mass nor electric charge yet. Even so, it is possible to know a priori what electric charge each particle will get within a given multiplet, by using a generalized Gell-Mann-Nishijima (GN) formula. The charges obtained in this way must be confirmed by the dynamics after the SSB is performed. 

The procedure to obtain electric charges is trivial when applied to fundamental representations and the GN formula is used directly, but this is not the case for fields that belong to higher dimension multiplets if we want to use the group generators written in the minimal dimension.

This happens in any gauge field theory with SSB. We will consider for illustration, extensions of the SM like $SU(3)_C\otimes SU(3)_L\otimes U(1)_X$, $SU(3)_C\otimes SU(2)_L\otimes SU(2)_R\otimes U(1)_Y$, and grand unified theories using $SU(5)$ as an example. In all these models a generalized GN formula $Q=T_3+Y/2$ is imposed as well as the electric charge conservation in all interactions, including the fermion-scalar (Yukawa) and fermion-vector boson interactions. The finite Abelian transformation is given by $U_Q=e^{ieQ}$, where $Q$ is a matrix satisfying the GN formula for each model.

In this way, we will show how to obtain the electric charge from representations of any dimension for the models mentioned above. Our method is based in considering the electric charge as a globally preserved quantum number when doing the SSB. It is very useful for representations of dimension $n\otimes n$ or $n\otimes n^*$ of groups $SU(n)$ or $SU(n)\otimes SU^\prime(n)$. We found, and this is the main result of ur paper, that there are two forms of how the electric charge operator acts on them, in units of the positron charge  $e$,
\begin{equation}
Q\Psi=[T,\Psi]+X\Psi,\quad Q\Phi=\{ T,\Phi\}+X\Phi,
\end{equation}
where $T$ denotes a linear combination of the diagonal generators of the group and $X$ the hypercharge which is defined differently for each model. In fact, $X$ is not necessarily proportional to the unity; for instance in the theory $SU(3)$ of mesons and baryons each $SU(2)$ subgroup has different strong hypercharge. 

The outline of this paper is as follows. In Sec. \ref{sec:su2}, we briefly review the method in models with $SU(2)$ and $SU(2)\otimes U(1)$ symmetries. The SM is an example of the latter type, in which considering besides the singlets and doublets, there are real and complex triplets represented as $2\times 2$ matrices. In Sec. III, we analyze the multiplets of $SU(2)$ lef-right symmetric model and found that the scalar bidoublets and triplets have similar charge operators like their SM counterparts despite they transform differently. In Sec. IV, we use the method again and found one novelty in the sextet charge operator: it appears as an anticommutator. In section V, we compute electric charges in $SU(3)$ left-right symmetric model where the scalar boson multiplets are the relevant cases. Finally, in Sec. VI we analyze the fermion sector of $SU(5)$ model.

\section{$\mathbf{SU(2)_V}$ and $\mathbf{SU(2)\otimes U(1)_Y}$ models}
\label{sec:su2}

Here we will consider models based on the global $SU(2)_V$ and local $SU(2)\otimes U(1)_Y$ symmetry. In the former case, hadrons (nucleons and pions) are considered while in the latter one, the cases of SM. Below, all electric charges are given in units of $e $.

\subsection{$SU(2)_V$: nucleons and mesons}
\label{subsec:nm}

Baryons and mesons may be classified under $SU(3)\supset SU(2)$ symmetry. With GN formula $Q=T_3+Y/2$, a doublet of baryons like nucleons $N\equiv \begin{pmatrix}p & n\end{pmatrix}^T$, or kaons $K\equiv \begin{pmatrix}K^+ & K^0\end{pmatrix}^T$, with $Y=B+S$, implies that $QN=\textrm{diag}(+1,0)$ ($B=1$ and $S=0$) and $QK=\textrm{diag}(+1,0)$ ($B=0$ and $S=1$).

Higher dimensional representation can be treated in similar way using the isospin generator in an appropriate dimensional representation. The triplet $T$ using  $T_3=\textrm{diag}(+1,0,-1)$ has $QT=\textrm{diag}(+1,0,-1)$ if $Y=0$ and $QT=\textrm{diag}(+2,+1,0)$ if $Y=2$.

However, we can obtain the same charge eigenvalues without resorting to higher generator representation. For example, a triplet can be written as a $2\times2$ matrix using the Pauli matrices $M_T=\sigma_a I_a/2$ and hence using $U_Q=\exp\left[ie(T_3+Y/2)\right]$, the transformation $U_QM_TU^\dagger_Q$ implies
\begin{equation}
 \delta_Q M_T\approx  QM_T=[Q,M_T]=\begin{pmatrix}
0t_1 & +1t_2\\
-1t_3 & 0t_1
\end{pmatrix},
\label{pions}
\end{equation}
at first order and $t_i$ are components of $M_T$.

In general we will omit the field components and write only the respective electric charge eigenvalues. It means that the matrix in Eq.~(\ref{pions}) is written just as
\begin{equation}
QM_T=\left( 
\begin{array}{cc}
0& +1\\
-1 & 0\\
\end{array}
\right).
\label{pion2}
\end{equation}

Hence, we see that it is not necessary to use the adjoint representation in order to find the electric charge of each triplet component.

This method can be used even if $SU(2)$ is a local symmetry and the triplet is formed with the gauge vector bosons because in the assignment of electric charge, only the global symmetry related with the $U_Q=e^{ie(T_3+Y/2)}$ is required.

A similar procedure can be applied for another multiplets. A quartet $\textbf{4}$, using $T_3=\frac{1}{2}\textrm{diag}(3,1,-1,-3)$ obtains $Q\textbf{4}=\textrm{diag}(+2,+1,0,-1)$ if $Y=1$, like the Delta baryon quartet, which can be defined from the known quark doublet $\begin{pmatrix}u & d\end{pmatrix}^T$ because in $SU(2)$ the quartet is obtained from the tensor product $\mathbf{2}\otimes \mathbf{2}\otimes \mathbf{2} = \mathbf{2}\oplus \mathbf{2}\oplus \mathbf{4}$. The symmetric part of $\textbf{2}\otimes \textbf{2}\otimes \textbf{2}$ is the quartet, $\textbf{4}$.

So, $(M_4)_{ijk}=\zeta_i\zeta_j\zeta_k$ with $\zeta$'s transforming as doublets $\zeta\rightarrow\zeta'\approx\left(1+i\frac{Y}{2}\right)\left(\mathbf{1}_2+iI_3\right)\zeta\approx U_Q\zeta$. Then, the charge operator is
\begin{equation}
\delta_Q M_4\approx QM_4=\acomm{M_4}{I_3}+\left(\zeta\zeta\left(\zeta I_3\right)^T+\frac{3Y}{2}\right)M_4.
\end{equation}

As $Y=\frac{1}{3}$ is the usual value for quark doublets, then 

\begin{equation}
QM_4=\begin{pmatrix}+2 & 0 \\ +1 & -1\end{pmatrix}.
\end{equation}

Because of the three indexes of $M_4$, we had to omit the repeated fields and achieved a $2\times 2$ matrix representations.

\subsection{The electroweak standard model}
\label{subsec:esm}

In spite of the SM is a well known case, including strong hadron interactions and the electroweak part, we consider this model in more detail  to illustrate our method and to apply it in the following gauge theories presented in this paper.

As before, electric charges are related by the GN formula  $Q=T_{3L}+\frac{Y}{2}$, being $T_{3L}$ the third generator of $SU(2)_L$ weak isospin, and $Y$ is the weak hypercharge. Besides, left-handed fermions and scalars, in their fundamental representations, transform like doublets.

For instance, consider a generic doublet $\mathcal{D}=(a\,b)^T\sim(\mathbf{2},Y)$. In this case
\begin{equation}\label{SM_Doublet_Transf}
\delta_Q \mathcal{D}\approx Q \mathcal{D}.
\end{equation}
As we said before, in the symmetric limit $a\leftrightarrow b$ because neither $a$ nor $b$ has mass or electric charge, but it is necessary to know \textit{a priori} which component is the one  which will gain mass, electric charge or a non-zero vacuum expectation value responsible of the SSB.

We can assume $U_Q=U(1)_YU_L$ with $U_L=e^{\alpha_aI_a/2}$ and $U(1)_Y=e^{if Y/2}$ under the conditions $\alpha_1(x)=\alpha_2(x)=0$ and $\alpha_3(x)=f(x)=\textrm{constant}$ can be assumed when SSB is taken into account. For the electric charge invariance, $\alpha_3\rightarrow e$, being $e$ the positron charge.
Then, we obtain Eq.~\eqref{SM_Doublet_Transf}  at first order,  and $Q=\textrm{diag}\left[\frac{1}{2}(1+Y,\,-1+Y)\right]$ according to GN formula. Choosing the hypercharges for left-handed doublets of quarks, leptons and Higgs as $Y(Q_L)=1/3,\,Y(L)=-1,\,Y(H)=+1$, their electric charge eigenvalues are $(+2/3,-1/3), (0,-1)$ and $(+1,0)$ in units of $e$.

On the other hand, SM right-handed fermions have singlet representations, i.e. $\alpha_j(x)=0$ for $j=1,2,3$. The transformation is
\begin{equation}
\delta_Qf_R\approx \frac{Y}{2}f_R,
\end{equation}
With the usual up-down quark hypercharge $Y(u_R)=\frac{4}{3}$ and $Y(d_R)=-\frac{2}{3}$, the known electric charges $Q(u_R)=+\frac{2}{3}$ and $Q(d_R)=-\frac{1}{3}$ are obtained. Similarly, right-handed charged leptons have $Y(l_R)=-2$ and $Q(l_R)=-1$. We see that there is no difference in the electric charge assignment if $SU(2)$ is a global, as in Subec.~\ref{subsec:nm}, or a local symmetry. Of course, in electroweak models after the SSB the electric charge conservation is related with  a local $U(1)$ symmetry. 

In the following we will write directly the global transformation related with the generator $Q$ in different models which is defined by a generalized GN formula.

A scalar, fermionic or even vector triplet $\begin{pmatrix}W_1 & W_2 & W_3\end{pmatrix}^T$ transforms under $SU(2)_L$ adjoint representation but is written as $W=\frac{\sigma_j}{2}W_j$ with $j=1,2,3$ in order to use the usual $2\times 2$ generators. We can write
\begin{equation}
W_{\mu}\equiv \frac{\sigma_j}{2}W_\mu^j=\frac{1}{2}\begin{pmatrix}
W_{\mu}^3 & W_{\mu}^1-iW_{\mu}^2 \\ W_{\mu}^1+iW_{\mu}^2 & -W_{\mu}^3,
\end{pmatrix},
\label{smws}
\end{equation}
and under the $U(1)_Q$ transforms as
\begin{equation}
\begin{split}
&W_{\mu}\rightarrow W_{\mu}^\prime=U_QW_{\mu}U_Q^{\dag}
\end{split}
\label{wssm}
\end{equation}

Then
\begin{equation}
\var_Q{W}_{\mu}=QW_{\mu}\approx [T^3_L,W_\mu]=\begin{pmatrix}
\hspace{0.2cm}0 & +1 \\ -1 & \hspace{0.2cm}0
\end{pmatrix},
\label{ecsmw}
\end{equation}
where the matrix entries are the electric charges of those fields defined in \eqref{smws}.

Also, the fourth vector boson $B_{\mu}$ that corresponds to the generator $Q=\frac{Y}{2}$ transforms as $B_{\mu}\rightarrow B_{\mu}'= B_{\mu}-\partial_{\mu}f(x)$ and since $f(x)=qe$, $B'_{\mu}\rightarrow B_{\mu}$ having no electric charge.
Then, the fields $Z_{\mu}$ and $A_{\mu}$ that are linear combinations of neutral $W_{\mu}^3$ and $B_{\mu}$ do not have electric charges neither.

There are another representation for triplets, like the one shown in Ref.~\cite{Cheng:1980qt}. Here, a complex scalar triplet $H=\begin{pmatrix}H_1 & H_2 & H_3\end{pmatrix}^t\sim\left(\textbf{3},2\right)$ had to be considered in order to perform the type-II seesaw model and 
\begin{equation}H=\frac{\sigma_j}{2}H_j=\begin{pmatrix}H_3 & h_1 \\ h_2 & H_3
\end{pmatrix}.
\label{ct}
\end{equation}

Unlike the real vector boson case in \eqref{smws}, $h_1\not =h_2^*$ because $H_j$'s are complex fields. In addition, the minus sign in the last $H_3$ entry has been omitted because electric charge is independent of it. Similarly as in the $W$'s case we obtain
\begin{equation}
QH=\comm{T_{3L}}{H}+\textbf{1}H=\begin{pmatrix}+1 & +2 \\ 0 & +1\end{pmatrix}.
\label{triplet1}
\end{equation}
Notice that $\epsilon H=\epsilon H_j\sigma_j/2$, where $\epsilon=i\sigma_2$ has the following electric charge eigenvalues
\begin{equation}
\left(\begin{array}{cc}
0 &+1\\
+1&+2 
\end{array}\right),
\label{triplet2}
\end{equation}
in appropriate form to generate a Majorana mass to neutrinos in terms like $\overline{L^c}\epsilon HL$~\cite{Cheng:1980qt}.

The electric charge assignment obtained so far has a relevant contribution because the values coincide with the results that were acquired from the usual procedure in which the electric charges (known and unknown) are confirmed after the SSB considering that charge conservation is additive in each interaction term.

However, what happens if we introduce fields that do not interact with the known quarks and leptons? Consider a symmetry $SU(2)^\prime_L\otimes U^\prime_Y$.

Let us suppose that the electric charge operator is defined with a generalized GN 
\begin{equation}
Q=aT_3+\frac{Y}{2}.
\end{equation}
Some possibilities are 
\begin{enumerate}[i)]
\item $a=2$ and $Y=0$. The electric charge eigenvalues are $(+1,-1)$. For instance a lepton doublet $\begin{pmatrix}E^+ & l^-\end{pmatrix}^T_L$ that couples to a real vector triplet which electric charge content is
\begin{equation}
Q\mathcal{T}\approx  [I_{3L},\mathcal{T}]=\begin{pmatrix}
\hspace{0.2cm}0 & +2 \\ -2  & \hspace{0.2cm}0 
\end{pmatrix},
\label{exotics}
\end{equation}
like $\mathcal{T}=\begin{pmatrix}U^{++} & W^{\prime0} & U^{--}\end{pmatrix}^T$.
\item $a=2$ and $Y=1$ obtain fields in doublets with electric charge eigenvalues $(+2,0)$.
\item The more exotic case when $a=1$ and $Y=-3$ include doublets with  electric charge $(-1,-2)$.  In this case it is possible to have doublets as $\begin{pmatrix}\vspace{+0.1cm}\chi^- & \chi^{--}\end{pmatrix}^T$ and they  couple with vector bosons like $\begin{pmatrix}V^+ & V^0 & V^-\end{pmatrix}^T$ if expressed as in \eqref{smws}.
\end{enumerate}

All these cases appear in the $SU(2)_L\times U(1)_Y$ projection of the minimal 3-3-1 model~\cite{Machado:2018sfh}. 

\section{Electric charges in SU(2) left-right symmetric models 
}
\label{sec:lr}

In this sort of models, parity is preserved at high energies and new right-handed fermionic doublets must be defined~\cite{Pati:1974yy}. In addition, the theory brings new right vector bosons and scalars multiplets different in mass from those of the SM.
The charge relation \cite{Marshak1980} is
\begin{equation}\label{221_Gell_Nish}
Q=T_{3L}+T_{3R}+\frac{B-L}{2},
\end{equation}
where $T_{3L}$ acts only over left multiplets and $T_{3R}$ over right ones. $B$ and $L$ are baryon and lepton number respectively.

\subsection{Fermions}
\label{subsec:lrf}

Quarks and leptons transform as left- and right-handed doublets for the three flavor families.
\begin{eqnarray}
&& Q_{iL}=(u_i ,\, d_i)^t\sim \left(\textbf{2}_L,\textbf{1}_R,+1/3\right),\quad
Q_{iR}=(u_i,\, d_i)^t\sim \left(\textbf{1}_L,\textbf{2}_R,+1/3\right),\nonumber \\&&
L_i=(\nu_i ,\, e_i)^t\sim \left(\textbf{2}_L,\textbf{1}_R,-1\right),\quad
R_i=(\nu_i ,\, e_i)^t_R\sim \left(\textbf{1}_L,\textbf{2}_R,-1\right),
\end{eqnarray}
where $i=1,2,3$. In this case we proceed as in the SM and the fermion doublets have the same electric charge assignment as those in the SM.

\subsection{Scalar Bosons}	
In this model the scalar multiplets can be represented as bidoublets and triplets \cite{Mohapatra1981}.
\begin{eqnarray}
&& \Phi=\begin{pmatrix}\phi_{11} & \phi_{12} \\ \phi_{21} & \phi_{22}\end{pmatrix}\sim \left(\textbf{2}_L,\textbf{2}_R^*,0\right),\quad  
\Delta_L=\begin{pmatrix}\delta_{11} & \delta_{12} \\ \delta_{21} & \delta_{22}
\end{pmatrix}\sim \left(\textbf{3}_L,\textbf{1}_R,+2\right), \nonumber \\&&
\Delta_R=\begin{pmatrix}\delta_{11} & \delta_{12} \\ \delta_{21} & \delta_{22}
\end{pmatrix}\sim \left(\textbf{1}_L,\textbf{3}_R,+2\right),
\end{eqnarray}
Bidoublets transform as
\begin{equation}
\Phi\rightarrow\Phi'=U_Q\Phi U_Q^{\dag}
\approx  \Phi+ie\left(T^3_L\Phi-\Phi T^3_R\right).
\end{equation}

So, their charge operator is $Q\Phi=[T_3,\Phi]$,  with $T_{3L}=T_{3R}=T_3$. Then,
\begin{equation}
Q\Phi=\begin{pmatrix}
0 & +1 \\ -1 & 0
\end{pmatrix}.
\end{equation}

Notice that $\phi_{11}$ and $\phi_{22}$ are both neutral scalars different from each other.

Besides, if we had a bifundamental multiplet $\Psi\sim \left(\textbf{2}_L,\textbf{2}_R,0\right)$ instead of $ \left(\textbf{2}_L,\textbf{2}_R^*,0\right)$, the charge operator would be 
\begin{equation}
\Psi\rightarrow\Psi=U_Q\Psi U_Q^T
\approx  \Phi+ie\left(T^3_L\Phi+\Phi T^3_R\right).
\end{equation}
or
\begin{equation}
Q\Psi=\acomm{T_3}{\Psi}=\begin{pmatrix}
+1 & 0 \\ 0 & -1
\end{pmatrix},
\end{equation}
but in this case, matrix entries are interchanged in each row ($\phi_{11}\rightarrow\phi'_{12}$ and $\phi_{21}\rightarrow\phi'_{22}$) due to the transformation nature of $SU(2)$ symmetry, resulting the same electric charges for the $\Phi$ and $\Psi$ components. 

For the triplets case, we have
\begin{equation}
\Delta'_{L,R}=U_Q\Delta_{L,R}U_Q^{\dag}
\approx \Delta_{L,R}+ie\left(T^3_{L,R}\Delta_{L,R}-\Delta_{L,R} T^3_{L,R}+\Delta_{L,R}\right).
\end{equation}

Its charge operator is $Q\Delta_{L,R}=\comm{T_3}{\Delta}_{L,R}+\Delta_{L,R}$ and the electric charges are
\begin{equation}
 Q\Delta_{L,R}=\begin{pmatrix}
+1 & +2 \\ 0 & +1
\end{pmatrix}.
\label{triplet2}
\end{equation}
As in Eq.~(\ref{triplet1}), a transformation by $\epsilon$ changes the position of the neutral component within the matrix and Majorana masses can be generated for left- and right-handed neutrinos.

\subsection{Vector Bosons}

The covariant derivatives for this group, have the same structure as those of the SM, but with additional vector bosons inherent to the $SU(2)_R$ adjoint representation. Then, the analysis is similar to the SM case in Eq.~(\ref{smws}) and the electric charge assignment is identical to that in the SM.

\section{Charges in the 3-3-1 models }
\label{sec:331}

These models are based on the $SU(3)_C\otimes SU(3)_L\otimes U(1)_N$ gauge symmetry. It is possible to choose a particle content representation in such a way that the anomaly cancellation accounts for three fermion families, unlike the SM, where the number of families is a free parameter (up to constraints coming from asymptotic freedom).

The electric charge relation is~\cite{Diaz:2004fs}
\begin{equation}\label{331_Gell_Nish}
Q=T_{3L}+\beta T_{8L}+N,
\end{equation}
where $T_{3L}$ and $T_{8L}$ are diagonal generators of $SU(3)$ group. Models with $\beta=-\sqrt3$ were considered in Refs.~\cite{Pisano1992,Frampton:1992wt,Foot:1992rh} while those with $\beta=-(1/\sqrt3)$ were considered in Refs.~\cite{Singer:1980sw,Montero:1992jk,Foot:1994ym}. The appropriate value for the charge $N$ depends on the value of $\beta$.

\subsection{Fermions}	
Quarks in $SU(3)_L$, transform like left-handed triplets for the three families. The particle content within these multiplets changes with each model version, but the charge operator is the same for all of them. 

The gauge transformations for left quarks are defined as
\begin{equation}\label{331_Quark_Transf}
\begin{split}
U_Q\mathcal{Q}_{1L}&\approx \left[\mathbf{1}+ie\left(T_{3L}+\beta T_{8L}\right)+ieN_1\mathbf{1}\right]\mathcal{Q}_{1L},\\
U^*_Q\mathcal{Q}_{aL}&\approx \left[\mathbf{1}-ie\left(T_{3L}+\beta T_{8L}\right)+ieN_2\mathbf{1}\right]\mathcal{Q}_{aL},
\end{split}
\end{equation}
where $T_j$ are group generators of $SU(3)_L$. 

According to equations Eqs.~\eqref{331_Gell_Nish} and \eqref{331_Quark_Transf}, the electric charge operators are
\begin{equation}\label{Charge_Op_331_Q}
\begin{split}
Q\mathcal{Q}_{1L}&=\begin{pmatrix}
\frac{1}{2}+\frac{\beta}{2\sqrt{3}}+N_1 & 0 & 0 \\ 0 & -\frac{1}{2}+\frac{\beta}{2\sqrt{3}}+N_1 & 0 \\ 0 & 0 & -\frac{\beta}{\sqrt{3}}+N_1
\end{pmatrix}\\[10pt]
Q\mathcal{Q}_{aL}&=\begin{pmatrix}
-\frac{1}{2}-\frac{\beta}{2\sqrt{3}}+N_2 & 0 & 0 \\ 0 & \frac{1}{2}-\frac{\beta}{2\sqrt{3}}+N_2 & 0 \\ 0 & 0 & \frac{\beta}{\sqrt{3}}+N_2
\end{pmatrix}.
\end{split}
\end{equation}

Thus,
\begin{equation*}
\begin{split}
Q(u_1)&=\frac{1}{2}+\frac{\beta}{2\sqrt{3}}+N_1,\hspace{0.5cm}Q(d_1)=-\frac{1}{2}+\frac{\beta}{2\sqrt{3}}+N_1,\hspace{0.5cm}Q(J_1)=-\frac{\beta}{\sqrt{3}}+N_1,\\
Q(d_a)&=-\frac{1}{2}-\frac{\beta}{2\sqrt{3}}+N_2,\hspace{0.5cm}Q(u_a)=\frac{1}{2}-\frac{\beta}{2\sqrt{3}}+N_2,\hspace{0.5cm}Q(J_a)=\frac{\beta}{\sqrt{3}}+N_2.
\end{split}
\end{equation*}

As usual, the hypercharges are chosen in such a way that the first two components of triplets have the known electrical charges of ``up'' and ``down'' quarks.

\begin{table}[h]
\caption{\label{Table03}Electric charge and hypercharge of Left quarks for different values of $\beta$.}
\begin{ruledtabular}
\begin{tabular}{c|cccc}
\vspace{0.1cm} &$\beta=-\sqrt{3}$ & $\beta=+\sqrt{3}$ & $\beta=-\frac{1}{\sqrt{3}}$ & $\beta=\frac{1}{\sqrt{3}}$\\
\colrule
$Q_{1L}$&$\left(\begin{array}{c} {\scriptstyle +2/3} \\ [-8pt]{\scriptstyle -1/3} \\[-8pt] {\scriptstyle +5/3}\end{array}\right),+\dfrac{2}{3}$ & $\left(\begin{array}{c} {\scriptstyle +2/3}\\[-8pt] {\scriptstyle -1/3} \\[-8pt] {\scriptstyle -4/3}\end{array}\right),-\dfrac{1}{3}$ & $\left(\begin{array}{c} {\scriptstyle +2/3} \\[-8pt] {\scriptstyle -1/3} \\[-8pt] {\scriptstyle +2/3}\end{array}\right),+\dfrac{1}{3}$ & $\left(\begin{array}{c} {\scriptstyle +2/3} \\[-8pt] {\scriptstyle -1/3} \\[-8pt] {\scriptstyle +1/3}\end{array}\right),0$\\  \colrule
$Q_{aL}$&$\left(\begin{array}{c} {\scriptstyle -1/3} \\[-8pt] {\scriptstyle +2/3} \\[-8pt] {\scriptstyle -4/3}\end{array}\right),-\dfrac{1}{3}$ & $\left(\begin{array}{c} {\scriptstyle -1/3} \\[-8pt] {\scriptstyle +2/3} \\[-8pt] {\scriptstyle +5/3}\end{array}\right),+\dfrac{2}{3}$ & $\left(\begin{array}{c} {\scriptstyle -1/3} \\[-8pt] {\scriptstyle +2/3} \\[-8pt] {\scriptstyle -1/3}\end{array}\right),0$ & $\left(\begin{array}{c} {\scriptstyle -1/3} \\[-8pt] {\scriptstyle +2/3} \\[-8pt] {\scriptstyle +2/3}\end{array}\right),+\dfrac{1}{3}$
\end{tabular}
\end{ruledtabular}
\end{table}

Leptons are triplets as well and the same charge operators in Eq.~\eqref{Charge_Op_331_Q} are applied on them. The left triplet hypercharge is chosen such that the SM doublets are recovered when the SSB is done. Just like in the quark case, the right counterparts are singlets and their hypercharges have the same values that their electric charges.
\begin{equation}
\begin{split}
L_i&=\begin{pmatrix}\nu_i & e_i & E_i\end{pmatrix}^T_L\sim \left(\textbf{3}_L,N_L\right)\\
\nu_{iR}&\sim\left(\textbf{1}_L,0\right),\,e_{iR}\sim\left(\textbf{1}_L,-1\right),\,E_{iR}\sim\left(\textbf{1}_L,N_R\right)
\end{split}
\end{equation}
%\vspace{-0.5cm}
The electric charge depends on the value of $\beta$ and $N_L$. When $\beta=-\sqrt{3}$, then $N_L=0$ and $E=E^+$ or $E=(l^c)_L$. In the later case there is not singlet charged leptons. The electric charge assignment for quarks in each model is shown in Table~\ref{Table03} and the same for leptons in Table~\ref{Table04}.

\begin{table}[h]
\caption{\label{Table04}Electric charge and hypercharge of Leptons for different values of $\beta$.}
\begin{ruledtabular}
\begin{tabular}{c|cccc}
\vspace{0.1cm} &$\beta=-\sqrt{3}$ & $\beta=+\sqrt{3}$ & $\beta=-\frac{1}{\sqrt{3}}$ & $\beta=\frac{1}{\sqrt{3}}$\\
\colrule
$L_i$&$\left(\begin{array}{c} 0 \\ [-8pt]-1 \\[-8pt] +1\end{array}\right),0$ & $\left(\begin{array}{c} 0\\[-8pt] -1 \\[-8pt] -2\end{array}\right),-1$ & $\left(\begin{array}{c} 0 \\[-8pt] -1 \\[-8pt] 0\end{array}\right),-\dfrac{1}{3}$ & $\left(\begin{array}{c} 0 \\[-8pt] -1 \\[-8pt] -1\end{array}\right),-\dfrac{2}{3}$\\ \colrule
$E_{iR}$&$+1$ & -2 & 0 & -1
\end{tabular}
\end{ruledtabular}
\end{table}

\subsection{Scalar Bosons}

All the 3-3-1 models need scalar triplets, say $\rho$, $\eta$ and $\chi$. Hence, we start by considering them using the electric charge operator in Eq.~\eqref{Charge_Op_331_Q}.  The electric charges for the different values of $\beta$ are given in Table III.

\begin{table}[h]
\caption{\label{Table05}Electric charge and hypercharge of Scalar triplets for different values of $\beta$.}
\begin{ruledtabular}
\begin{tabular}{c|cccc}
\vspace{0.1cm} &$\beta=-\sqrt{3}$ & $\beta=+\sqrt{3}$ & $\beta=-\frac{1}{\sqrt{3}}$ & $\beta=\frac{1}{\sqrt{3}}$\\
\colrule
\hspace{0.5cm}$\eta$&$\left(\begin{array}{c} 0 \\ [-8pt]-1 \\[-8pt] +1\end{array}\right),0$ & $\left(\begin{array}{c} 0\\[-8pt] -1 \\[-8pt] -2\end{array}\right),-1$ & $\left(\begin{array}{c} 0 \\[-8pt] -1 \\[-8pt] 0\end{array}\right),-\dfrac{1}{3}$ & $\left(\begin{array}{c} 0 \\[-8pt] -1 \\[-8pt] -1\end{array}\right),-\dfrac{2}{3}$\\ \colrule
\hspace{0.5cm}$\rho$&$\left(\begin{array}{c} +1 \\ [-8pt]0 \\[-8pt] +2\end{array}\right),+1$ & $\left(\begin{array}{c} +1\\[-8pt] 0 \\[-8pt] -1\end{array}\right),0$ & $\left(\begin{array}{c} +1 \\[-8pt] 0 \\[-8pt] +1\end{array}\right),+\dfrac{2}{3}$ & $\left(\begin{array}{c} +1 \\[-8pt] 0 \\[-8pt] 0\end{array}\right),+\dfrac{1}{3}$\\ \colrule
\hspace{0.5cm}$\chi$&$\left(\begin{array}{c} -1 \\ [-8pt]-2 \\[-8pt] 0\end{array}\right),-1$ & $\left(\begin{array}{c} +2\\[-8pt] +1 \\[-8pt] 0\end{array}\right),+1$ & $\left(\begin{array}{c} 0 \\[-8pt] -1 \\[-8pt] 0\end{array}\right),-\dfrac{1}{3}$ & $\left(\begin{array}{c} +1 \\[-8pt] 0 \\[-8pt] 0\end{array}\right),+\dfrac{1}{3}$
\end{tabular}
\end{ruledtabular}
\end{table}

It is also possible to introduce a scalar sextet, $S$ \cite{Machado:2018sfh,Montero:2001ji}. For any value of $\beta$, the mentioned sextet is defined as
\begin{equation}\label{331_Scalars1}
S=\begin{pmatrix}
\sigma_1 & \sigma_2 & \sigma_3 \\
\sigma_2 & \sigma_4 & \sigma_5 \\
\sigma_3 & \sigma_5 & \sigma_6
\end{pmatrix}\hspace{-0.15cm}\sim(\textbf{6}_L,N).
\end{equation}

Since $3\times 3=3_a^*+6_s$~\cite{Slansky1981}, these tensors can be written as  $\epsilon_{abc}\epsilon^{cde}u_dv_e+\frac{1}{2}\left(u_av_b+u_bv_a\right)$. Then, the second addend is a sextet that can be expressed as $S=\frac{1}{2}\left(\Lambda+\Lambda^T\right)$, where $\Lambda=uv^t$ with $u$ and $v$ are triplet fundamental representations. Tensor $\Lambda_{ab}$  transforms as
\begin{equation}
\label{331_S_Charge_Op_deduction}
\Lambda^\prime=(U_Qu)(v^tU^T_Q) \approx\Lambda+ie\acomm{\Lambda}{T^{3L}+\beta T^{8L}}+ieN\Lambda,
\end{equation}
where $N=N_1+N_2$.
As $\Lambda^T$ transforms in the same way, therefore
\begin{equation}
QS=\acomm{T^3_L+\beta T^8_L}{S}+NS.
\label{qs}
\end{equation}

Explicitly,
\begin{equation}\label{Charge_Op_331_S_beta}
QS=\begin{pmatrix}
1+\frac{\beta}{\sqrt{3}}+N & \frac{\beta}{\sqrt{3}}+N & \frac{1}{2}\left(1-\frac{\beta}{\sqrt{3}}+2N\right)\\
\frac{\beta}{\sqrt{3}}+N & -1+\frac{\beta}{\sqrt{3}}+N & -\frac{1}{2}\left(1+\frac{\beta}{\sqrt{3}}-2N\right)\\
\frac{1}{2}\left(1-\frac{\beta}{\sqrt{3}}+2N\right) & -\frac{1}{2}\left(1+\frac{\beta}{\sqrt{3}}-2N\right) & -\frac{2\beta}{\sqrt{3}}+N
\end{pmatrix}_{\hspace{-0.1cm}S}
\end{equation}

The sextet acquires hypercharge from the leptons in Table \eqref{Table04}, and for the possible values of $\beta$,
\begin{equation}\label{Charge_Op_331_S}
\begin{split}
&\begin{array}{l}\beta=-\sqrt{3}\\N=0\end{array}: QS=\begin{pmatrix}
0 & -1 & +1 \\ -1 & -2 & 0 \\ +1 & 0 & +2\end{pmatrix}_{\hspace{-0.1cm}S},\hspace{1cm}
\begin{array}{l}\beta=+\sqrt{3}\\N=-2\end{array}: QS=\begin{pmatrix} 0 & -1 & -2 \\ -1 & -2 & -3 \\ -2 & -3 & -4\end{pmatrix}_{\hspace{-0.1cm}S},\\
&\begin{array}{l}\beta=-\frac{1}{\sqrt{3}}\\N=-\frac{2}{3}\end{array}: QS=\begin{pmatrix}0 & -1 & 0 \\ -1 & -2 & -1 \\ 0 & -1 & 0\end{pmatrix}_{\hspace{-0.1cm}S},\hspace{1cm} \begin{array}{l}\beta=+\frac{1}{\sqrt{3}}\\N=-\frac{4}{3}\end{array}: QS=\begin{pmatrix}
0 & -1 & -1 \\ -1 & -2 & -2 \\ -1 & -2 & -2\end{pmatrix}_{\hspace{-0.1cm}S}.
\end{split}
\end{equation}

Notice that for each $\beta$ there is a different number of neutral scalars (Higgs) responsible for assigning Majorana mass terms to neutrinos, which means that the couplings (and the whole theory) depend on $\beta$.

\subsection{Vector Bosons}
As in the $SU(2)_L$ case, the eight vector bosons for $SU(3)_L$ are arranged in the matrix $W_{\mu}=\frac{\lambda_j}{2}W_{\mu j},\,j=1,\ldots,8$, where $\lambda_j$ are the Gell-Mann matrices. Explicitly we have
\begin{equation}
W_{\mu}=\frac{1}{2}\begin{pmatrix}
W_{\mu}^3+\frac{W_{\mu}^8}{\sqrt{3}} & \sqrt{2}W_{\mu}^{\dag} & \sqrt{2}V_{\mu} \\ \sqrt{2}W_{\mu} & -W_{\mu}^3+\frac{W_{\mu}^8}{\sqrt{3}} & \sqrt{2}U_{\mu} \\ \sqrt{2}V_{\mu}^{\dag} & \sqrt{2}U_{\mu}^{\dag} & -\frac{2}{\sqrt{3}}W_{\mu}^8
\end{pmatrix}
\label{331vb}
\end{equation}

In this case we have
\begin{equation}
QW_{\mu}=\comm{T^3_L+\beta T^8_L}{W_{\mu}}=\begin{pmatrix}
0 & +1 & q(V) \\ -1 & 0 & q(U) \\ -q(V) & -q(U) & 0
\end{pmatrix},
\label{cw331}
\end{equation}
where $q(V)=\frac{1}{2}\left(1+\frac{3\beta}{\sqrt{3}}\right)$ and $q(U)=\frac{1}{2}\left(-1+\frac{3\beta}{\sqrt{3}}\right)$. Again, the matrix entries are the electric charges of \eqref{331vb}.

For the different values of $\beta$ we have: 
\begin{equation*}
\begin{split}
&\beta=-\sqrt{3}:QW_{\mu}=\begin{pmatrix} 0 & +1 & -1 \\ -1 & 0 & -2 \\ +1 & +2 & 0 \end{pmatrix},
\hspace{0.5cm}\beta=+\sqrt{3}:QW_{\mu}=\begin{pmatrix} 0 & +1 & +2 \\ -1 & 0 & +1 \\ -2 & -1 & 0 \end{pmatrix},\\
&\beta=-\frac{1}{\sqrt{3}}:QW_{\mu}=\begin{pmatrix}
0 & +1 & 0 \\ -1 & 0 & -1 \\ 0 & +1 & 0 \end{pmatrix},
\hspace{0.5cm}\beta=+\frac{1}{\sqrt{3}}:QW_{\mu}=\begin{pmatrix} 0 & +1 & +1 \\ -1 & 0 & 0 \\ -1 & 0 & 0\end{pmatrix}.
\end{split}
\end{equation*}

From these equations and the diagonal of \eqref{331vb}, it can be noted that there are two neutral scalars, $W^3_{\mu}$ and $W^8_{\mu}$, in all four possibilities of the $\beta$ parameter. Moreover, there are new charged vector bosons and in the case $\beta=\pm(1/\sqrt3)$ there is one additional neutral complex field. This opens up a large number of interactions that describe new physics.

\section{Electric charges in the SU(3) left-right symmetric models }
\label{sec:su3lr}

This model combines the benefits of the previous two ones; left-right models preserves parity symmetry and 331 models explain why only three families are observed.

The charge relation is
\begin{equation}\label{3331_Gell_Nish}
Q=T_{3L}+T_{3R}+\beta\left(T_{8L}+T_{8R}\right)+X.
\end{equation}

The content particles according to Ref.~\cite{Reig:2016tuk,Franco:2016hip} will be considered, but for a general $\beta$.

\subsection{Fermions}
Quarks and leptons are left- and right-handed triplets (or antitriplets) like those of 331 case, and their charge operators are defined in equations \eqref{Charge_Op_331_Q}.

These six multiplets duplicate particles of the 331 chiral model of the chiral 331 model considering the right fermions additionally.

\subsection{Scalar bosons}

Four scalar multiplets are proposed \cite{Reig:2016tuk,Franco:2016hip}: one bitriplet $\Phi$, one bifundamental $P$ and two symmetric  sextets $S_{L,R}$.
\begin{equation}
\begin{split}
\Phi&=\begin{pmatrix}
\eta_1 & \rho_1 & \chi_1 \\
\eta_2 & \rho_2 & \chi_2 \\
\eta_3 & \rho_3 &\chi_3
\end{pmatrix}\sim (\mathbf{1},\mathbf{3_L},\mathbf{3_R^*},0),\;
P=\begin{pmatrix}
p_{11} & p_{12} & p_{13} \\
p_{21} & p_{22} & p_{23}\\
p_{31} & p_{32} & p_{33}
\end{pmatrix}\sim (\mathbf{1},\mathbf{3_L},\mathbf{3_R},X_P),\\
S_L&=\begin{pmatrix}
\sigma_1 & \sigma_2 & \sigma_3 \\
\sigma_2 & \sigma_4 & \sigma_5 \\
\sigma_3 & \sigma_5 & \sigma_6
\end{pmatrix}_L\sim (\mathbf{1},\mathbf{6_L},\mathbf{1_R},X_S),\;
S_R=\begin{pmatrix}
\sigma_1 & \sigma_2 & \sigma_3 \\
\sigma_2 & \sigma_4 & \sigma_5 \\
\sigma_3 & \sigma_5 & \sigma_6
\end{pmatrix}_R\sim (\mathbf{1},\mathbf{1_L},\mathbf{6_R},X_S).
\end{split}
\end{equation}
The bitriplet can be built from a left triplet and right antitriplet, $\Phi=v_Lv_R^{\dag}$, then the transformation is
\begin{equation}
\Phi\rightarrow\Phi'=\left(U_Qv_L\right)\left(U_Qv_R\right)^{\dag}=U_Q\Phi U^{\dag}_Q\approx \Phi+ie\comm{T_3+\beta T_8}{\Phi},
\end{equation}
where the usual values of $\alpha_j(x)$ and $f(x)$ have been assumed. The bitriplet has zero hypercharge as a consequence of its transformation $U_L\Phi U_R^\dag$, because the hypercharges of $v_L$ and $v_R^\dag$ have opposite signs, unlike the sextet in \eqref{qs} where the hypercharge triplets had to be added.

The charge operator for the bitriplet is $Q\Phi=\comm{T_3+\beta T_8}{\Phi}$,
\begin{equation}
\begin{split}
&Q\Phi=\begin{pmatrix}
0 & +1 & \frac{1}{2}\left(1+\frac{3\beta}{\sqrt{3}}\right) \\
-1 & 0 & \frac{1}{2}\left(-1+\frac{3\beta}{\sqrt{3}}\right) \\
-\frac{1}{2}\left(1+\frac{3\beta}{\sqrt{3}}\right) & \frac{1}{2}\left(1-\frac{3\beta}{\sqrt{3}}\right) & 0
\end{pmatrix}.
\end{split}
\end{equation}

For the possible values of $\beta$,
\begin{equation}
\begin{split}
\beta&=-\sqrt{3}: Q\Phi=\begin{pmatrix}
0 & +1 & -1 \\
-1 & 0 & -2 \\
+1 & +2 & 0
\end{pmatrix},\hspace{0.5cm}
\beta=+\sqrt{3}: Q\Phi=\begin{pmatrix}
0 & +1 & +2 \\
-1 & 0 & +1 \\
-2 & -1 & 0
\end{pmatrix},\\
\beta&=-\frac{1}{\sqrt{3}}: Q\Phi=\begin{pmatrix}
0 & +1 & 0 \\
-1 & 0 & -1 \\
0 & +1 & 0
\end{pmatrix},\hspace{0.5cm}
\beta=\frac{1}{\sqrt{3}}: Q\Phi=\begin{pmatrix}
0 & +1 & +1 \\
-1 & 0 & 0 \\
-1 & 0 & 0
\end{pmatrix}.
\end{split}
\end{equation}

This result is consistent with charges in table \eqref{Table05} and the fact that bitriplet is made up of $\Phi=\begin{pmatrix}\eta &\rho & \chi\end{pmatrix}^T$ for any $\beta$.

The bifundamental multiplet can be built from a left and right triplet, $\Phi=w_Lw_R^T$ with hypercharge $X'$, whereupon the transformation is 
\begin{equation}
P\rightarrow P'=(U_Qw_L)(U_Qw_R)^T=U_QPU_Q^T\approx P+ie\acomm{T^3+\beta T^8}{P}+ieXP,
\end{equation}
where $X=2X'$. The charge operator for $P$ is $QP=\acomm{T^3+\beta T^8}{P}+XP$
\begin{equation}
\begin{split}
QP=&=\begin{pmatrix}
\left(1+\frac{\beta}{\sqrt{3}}+X\right) & \left(\frac{\beta}{\sqrt{3}}+X\right) & \frac{1}{2}\left(1-\frac{\beta}{\sqrt{3}}+2X\right) \\
\left(\frac{\beta}{\sqrt{3}}+X\right) & \left(-1+\frac{\beta}{\sqrt{3}}+X\right) & -\frac{1}{2}\left(1+\frac{\beta}{\sqrt{3}}-2X\right) \\
\frac{1}{2}\left(1-\frac{\beta}{\sqrt{3}}+2X\right) & -\frac{1}{2}\left(1+\frac{\beta}{\sqrt{3}}-2X\right) & \left(-\frac{2\beta}{\sqrt{3}}+X\right)\end{pmatrix},
\end{split}
\end{equation}
and this result shows that $P$ may be considered as a symmetric multiplet.

The bifundamental multiplet is designed to be coupled with quarks, e.g. $\overline{Q}_{1L}PQ_{aR}$. Therefore, the hypercharge assignment depends on $\beta$.
\begin{equation*}
\begin{split}
 &\begin{array}{l} \beta=-\sqrt{3} \\ X=+1 \end{array}: QP=\begin{pmatrix}+1 & 0 & +2 \\ 0 & -1 & +1 \\ +2 & +1 & +3 \end{pmatrix}\hspace{0.5cm} 
\begin{array}{l} \beta=+\sqrt{3} \\ X=-1 \end{array}:QP=\begin{pmatrix}+1 & 0 & -1 \\0 & -1 & -2 \\-1 & -2 & -3\end{pmatrix},\\
 &\begin{array}{l} \beta=-\frac{1}{\sqrt{3}} \\ X=+\frac{1}{3}\end{array}:QP=\begin{pmatrix}+1 & 0 & +1 \\0 & -1 & 0 \\+1 & 0 & +1
\end{pmatrix},\hspace{0.5cm} 
\begin{array}{l}\beta=\frac{1}{\sqrt{3}} \\ X=-\frac{1}{3}\end{array}: QP=\begin{pmatrix} +1 & 0 & 0 \\0 & -1 & -1 \\0 & -1 & -1\end{pmatrix}.
\end{split}
\end{equation*}

Finally, the two sextets $S_{L,R}$ receive the same treatment as the sextet of model 331 and the electric charges of their components are the same as equation \eqref{Charge_Op_331_S} for the left- and right- cases.

\subsection{Vector Bosons}
Like in $SU(2)$-LR model that has the same vector boson matrix as the SM, but duplicated for the right bosons, the same happens for $SU(3)$-LR model. That is, the $W_{\mu}^{L,R}$ matrix has the same shape of the chiral model version in Eq. \eqref{331vb}, but for two cases: L and R.

Therefore,
\begin{equation}
QW_{\mu}^{L,R}=[T^3+\beta T^8,W^{L,R}_\mu],
\end{equation}
and the electric charge content is the same as Eq.~\eqref{cw331}.

\section{SU(5)}
\label{sec:su5}

In $SU(5)$ theory~\cite{Georgi:1974sy} the electric charge assignment is given by $Q=T_3+\frac{Y}{2}$ as usual but now with $(T_3)_{5\times 5}= (1/2)(\alpha T^4_4-\beta T^5_5)$ where the $T^a_b$ are the $SU(5)$ generators in the notation of Ref.~\cite{Langacker:1980js}.

For $\alpha=-\frac{\sqrt{6}}{2}$ and $\beta=-\frac{\sqrt{10}}{2}$, the matrix $T_3=\text{diag}\left(0,0,0,\frac{1}{2},-\frac{1}{2}\right)$. Besides, the hypercharge matrix is $Y=\text{diag}\left(y_i,y_j\right),\; i=1,2,3,\;j=1,2$; where $y_i$ is the hypercharge for $SU(3)$ color-triplet fermion and $y_j$ is the $SU(2)$ doublet hypercharge. Then the fermion/antifermion representation of charge operator is
\begin{equation}\label{SU(5)_Gell_Nish}
Q\Psi=\text{diag}\left[\frac{1}{2}\left(y_i,y+1,y-1\right)\right],\;Q^*\Psi^*=\text{diag}\left[\frac{1}{2}\left(y^*_i,y^*-1,y^*+1\right)\right],
\end{equation}
when $\Psi_5$ is a $\textbf{5}$ and $\Psi^*_5$ is an $\textbf{5}^*$.

Each fermion family of fifteen fields, may be reduced in a $5+10$ dimensional representation. Right-handed neutrinos are assumed to be singlet leptons,

Be $\Psi^*_{5L}$ a left-handed antiquintet that can be expressed as a sum of a SM color triplet and a electroweak doublet, $\mathbf{5^*}=\left(\mathbf{3^*},\mathbf{1},+\frac{2}{3}\right)+\left(\mathbf{1},\mathbf{2^*},-1\right)$,
which means that the weak hypercharges are $y_i=\frac{2}{3}$ and $y=-1$ in \eqref{SU(5)_Gell_Nish}.
\begin{equation}
Q\Psi_L=\text{diag}\left(\frac{1}{3},\frac{1}{3},\frac{1}{3},-1,0\right)\Psi_L.
\end{equation}

It is consistent with the usual representation $\Psi_{5L}=\begin{pmatrix}d_i^c & e^- &-\nu_e\end{pmatrix}^T_L$, with $i=1,2,3$ color index.
Another choice $\mathbf{5^*}=\left(\mathbf{3^*},\mathbf{1},-\frac{4}{3}\right)+\left(\mathbf{1},\mathbf{2^*},-1\right)$, is consistent with $\Psi_{5L}=\begin{pmatrix}u_i^c & e^- &-\nu_e\end{pmatrix}^T_L$.

In the same way, $\Psi_{5R}$ is a right-handed quintet that transforms as $\mathbf{5}=\left(\mathbf{3},\mathbf{1},-\frac{2}{3}\right)+\left(\mathbf{1},\mathbf{2},+1\right)$. The charge operator is

\begin{equation}
Q\Psi_R=\text{diag}\left(-\frac{1}{3},-\frac{1}{3},-\frac{1}{3},1,0\right),
\end{equation}
which is consistent with the representation $\Psi_{5R}=\begin{pmatrix}d_i & e^+ &-\nu^c_e\end{pmatrix}^T_R$. The decomposition $\mathbf{5}=\left(\mathbf{3},\mathbf{1},\frac{4}{3}\right)+\left(\mathbf{1},\mathbf{2},+1\right)$ corresponds to the representation $\Psi_{5R}=\begin{pmatrix}u_i & e^+ &-\nu^c_e\end{pmatrix}^T_R$.

On the other hand, this theory considers a decuplet that comes from $5\otimes 5=10_A\oplus15_S$ \cite{Langacker:1980js}. Then, a generalized antisymmetric decuplet may be represented as

\begin{equation}
\Psi_{10}=\begin{pmatrix}
0 & \varphi_1 & \varphi_2 & \varphi_3 & \varphi_4 \\ -\varphi_1 & 0 & \varphi_5 & \varphi_6 & \varphi_7 \\ -\varphi_2 & -\varphi_5 & 0 & \varphi_8 & \varphi_9 \\ -\varphi_3 & -\varphi_6 & -\varphi_8 & 0 & \varphi_{10} \\ -\varphi_4 & -\varphi_7 & -\varphi_9 & -\varphi_{10} & 0
\end{pmatrix},
\end{equation}
which can be defined as $\Psi_{10}^{ab}=\frac{1}{2}\left(\Psi_5^a\Psi_5^b-\Psi_5^b\Psi_5^a\right)$. For the left-handed decuplet, $\Psi_5\rightarrow \Psi_{5L}$ which transforms as an anti-quintet.

In the same way as in the $SU(3)$ sextet case, be $\Omega^{ab}_L=\Psi_{5L}^a\Psi_{5L}^b$ that transforms

\begin{equation}
\Omega_L \rightarrow \left(U_L\Psi_{5L}\right)\left(U_L\Psi_{5L}\right)\approx \Omega_+ie\acomm{\Omega_L}{-T^3+\frac{Y^*}{2}}.
\end{equation}

As $\Omega_L^T$ has the same transformation then 

\begin{equation}
Q\Psi_{10L}=\acomm{\Omega_L}{-T^3+\frac{Y^*}{2}}=\begin{pmatrix}
0 & -2/3 & -2/3 & 2/3 & -1/3 \\ -2/3 & 0 & -2/3 & 2/3 & -1/3 \\ -2/3 & -2/3 & 0 & 2/3 & -1/3 \\  2/3 & 2/3 & 2/3 & 0 & 1 \\ -1/3 & -1/3 & -1/3 & 1 & 0 
\end{pmatrix},
\end{equation}
which corresponds to the usual left-handed decuplet \cite{Langacker:1980js}.

For the right-handed decuplet

\begin{equation}
Q\Psi_{10R}=\acomm{\Omega_L}{T^3+\frac{Y}{2}}=\begin{pmatrix}
0 & 2/3 & 2/3 & -2/3 & 1/3 \\ 2/3 & 0 & 2/3 & -2/3 & 1/3 \\ 2/3 & 2/3 & 0 & -2/3 & 1/3 \\  -2/3 & -2/3 & -2/3 & 0 & -1 \\ 1/3 & 1/3 & 1/3 & -1 & 0 
\end{pmatrix}.
\end{equation}

\section{Conclusions}
\label{sec:con}

Here we have proposed a method to obtain the electric charges of the particles with global or local  gauge theories. In the latter case the electric charge assignment is done  before the SSB. It has been tested in five different theories and has been successful in all cases. The validity of the method is done for particles whose electric charges were already known, generating confidence in the results for charges of exotic particles.

On the other hand, the electric charge eigenstates obtained give us some properties of scalar multiplets (especially when these are matrix representations $2\otimes 2$, $3\otimes 3$, etc.) when developing new theories. For example, the method developed for the 331 sextet case told to us how it transforms as well as for $\Phi$ and $P$ scalar multiplets of the $SU(3)$-LR model.

In the case of vector bosons, there is an additional contribution: it was demonstrated that their electric charge can be computed in their fundamental representation or in the adjoint one. However, it must be said that the electric charges found are those of the gauge eigenstates and not those of mass eigenstates. Despite this, the physical states must preserve the electric charge obtained.

\section*{Acknowledgements}
ECR thanks to CONCYTEC for financial support  and to the IFT-UNESP for the kind hospitality where part of this work was done.
VP would like to thanks for partial financial support to CNPq and FAPESP under the funding Grant No. 2014/19164-6 and last, but not least, to the Faculty of Sciences of the Universidad Nacional de Ingenierıía (UNI) for the kind hospitality. OPR would would like to thank FAPESP grant 2016/01343-7 for funding my visit to ICTP-SAIFR from 01-08/12/2019 where part of this work was done and last, but not least, to IFT-UNESP for the kind hospitality.

\end{document}